\documentclass[twocolumn,showpacs,preprintnumbers,amsmath,amssymb,superscriptaddress]{revtex4}

\usepackage{graphicx}
\usepackage{dcolumn}
\usepackage{bm}

\begin{document}

\preprint{arXiv preprint}

\title{Emission spectrum of a dressed exciton-biexciton complex in a semiconductor quantum dot}

\author{Andreas Muller}
 \email{andreas.muller@nist.gov}
\affiliation{%
Joint Quantum Institute, National Institute of Standards and Technology and University of Maryland, Gaithersburg, MD 20899}
\author{Wei Fang}%
\affiliation{%
Joint Quantum Institute, National Institute of Standards and Technology and University of Maryland, Gaithersburg, MD 20899}
\author{John Lawall}
\affiliation{
Atomic Physics Division, National Institute of Standards and Technology, Gaithersburg, MD 20899
}%
\author{Glenn S. Solomon}%
\affiliation{%
Joint Quantum Institute, National Institute of Standards and Technology and University of Maryland, Gaithersburg, MD 20899}
\affiliation{
Atomic Physics Division, National Institute of Standards and Technology, Gaithersburg, MD 20899
}

\date{\today}

\begin{abstract}
The photoluminescence spectrum of a single quantum dot was recorded as a secondary resonant laser optically dressed either the vacuum-to-exciton or the exciton-to-biexciton transitions. High-resolution polarization-resolved measurements using a scanning Fabry-P\'erot interferometer reveal splittings of the linearly-polarized fine-structure states that are non-degenerate in an asymmetric quantum dot. These splittings manifest as either triplets or doublets and depend sensitively on laser intensity and detuning. Our approach realizes complete resonant control of a multi-excitonic system in {\it emission}, which can be either pulsed or continuous-wave, and offers direct access to the emitted photons.
\end{abstract}

\maketitle

Semiconductor quantum dots (QDs) \cite{bayer2000hse} have attracted steady attention due to narrow optical linewidths \cite{borri2001udt}, large and stable photon flux, and the robustness and scalability typically associated with well-developed semiconductor technology. This makes them promising for nanoscale quantum optics and applications such as quantum key distribution.

Recent experiments with coherently-controlled QDs have reinforced the feasibility of such goals, and established an ever-more striking resemblance to single atoms \cite{xu2007cos, jundt2008ode, muller2007rfc, wrigge2008ecp}. Of particular interest here is the manipulation of QD states using strong continuous-wave (cw) lasers, which gives rise to hybrid matter-field systems often described in the ``dressed-state'' picture \cite{cohentannoudji}. For example, the Autler-Townes effect in the fine-structure of a single QD exciton, as well as non-linear absorption due to interference effects between a strong pump and a weak probe laser have been reported in neutral \cite{xu2007cos} and charged \cite{kroner2008rsa} conditions. Dressing of an exciton ground state via a biexciton transition has also been realized using Stark shift modulation absorption spectroscopy \cite{jundt2008ode}, demonstrating optical polarization and the AC Stark effect. Furthermore, resonance fluorescence has been obtained from QDs \cite{muller2007rfc, muller2007cds} and single molecules \cite{wrigge2008ecp}, and the fluorescent Mollow triplet was demonstrated. These results build on substantial prior work on ultra-fast manipulation \cite{unold2004ose}, quantum interference \cite{bonadeo1998coc}, and Rabi oscillations \cite{stievater2001roe, htoon2002iro, kamada2001ero, zrenner2002cpt} in single QD systems.

Despite considerable improvement in optical control, however, most experiments are still restricted to absorption measurements and the study of the photon emission has been limited. Because of large QD dipole moments ($\approx$10 Debye), emission at optical frequencies can be substantial allowing experiments on the created photonic states.  Single \cite{michler2000qds, santori2001tsp} and entangled \cite{akopian2006epp, stevenson2006sst} photon sources, the Purcell effect \cite{deppe1999ese, gerard1998ese}, and vacuum Rabi oscillations \cite{yoshie2004vrs, reithmaier2004scs, peter2005eps, park2006cqd}, have all been demonstrated, and efforts are underway towards more advanced effects such as coherent population trapping \cite{santori2006cpt} and qubit-photon interfaces \cite{yao2005tcs, kiraz2004qds}. In this context it is important to achieve coherent control in emission measurements, for example to take advantage of the biexciton radiative cascade to produce polarization entangled photon pairs in a deterministic fashion \cite{jundt2008ode}. While absorption approaches are inappropriate, resonance fluorescence is also insufficient since both exciton and biexciton states need to be excited.

We report here on the full dressing of an exciton-biexciton complex in a single semiconductor QD in an experiment revealing the unique features of a dressed four-level system in emission. A pump laser at a frequency above the GaAs band-edge is used to generate photoluminescence (PL) from a single QD that is simultaneously dressed by a near resonant control laser. This results in a unique, and to our knowledge unexplored approach, in which the above-band laser serves primarily to populate the levels that have been tailored by the control laser. We show that, unlike in pump-probe techniques, the two lasers are largely independent in the sense that the pump laser frequency or amplitude (that can be large) does not directly affect the shape of the emission spectrum. In particular, when applying a pulsed pump laser, we can obtain a pulsed spectrum that remains dressed by the control laser.

\begin{figure}[t!]
\includegraphics[width=3.5in]{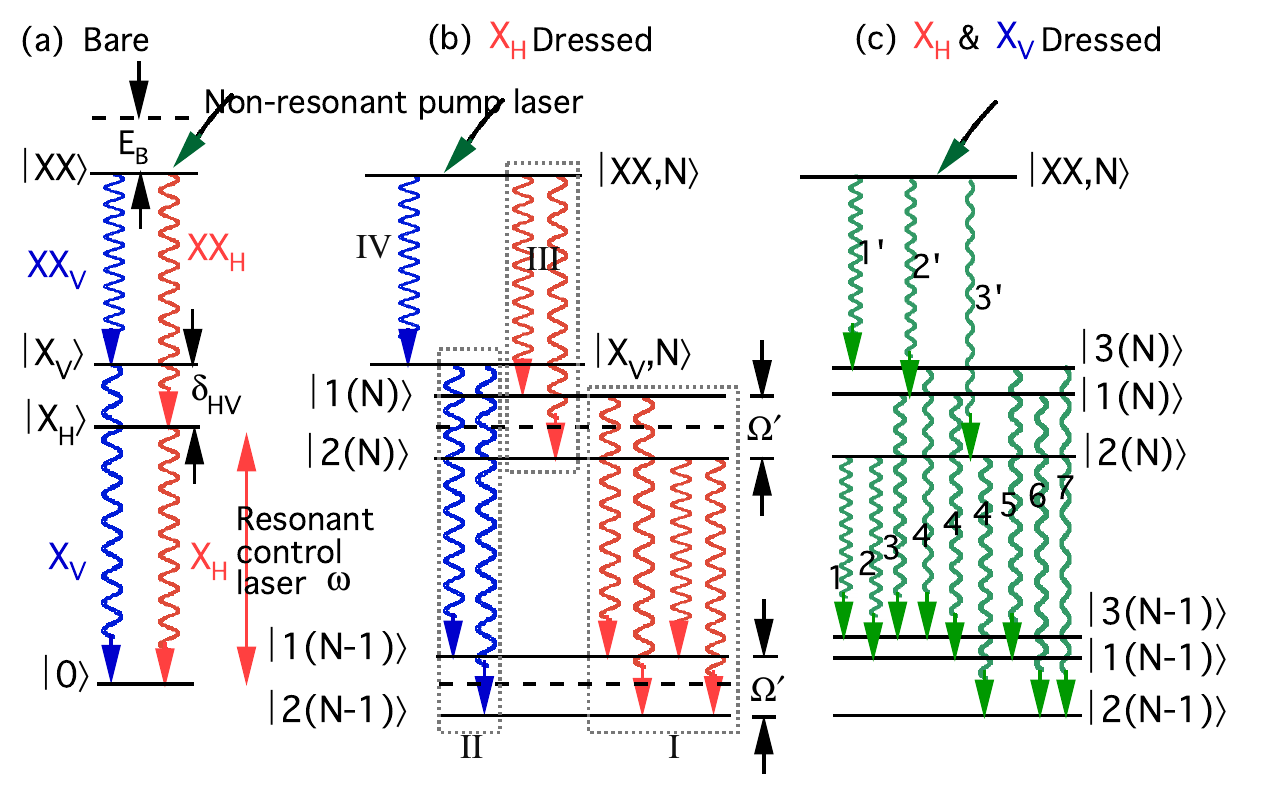}
\caption{\label{fig1} (color online). Energy schematic and the allowed transitions in the bare (a), $\rm X_H$ dressed (b), and both $\rm X_{H}$ and $\rm X_{V}$ dressed (c) exciton-biexciton complex. Dashed box I contains the triplet emission while II and III are Autler-Townes doublets. IV is unchanged. In (b) and (c) only the relevant levels of the N and N-1 manifolds are shown.}
\end{figure}

The investigated sample contains a dilute ensemble of self-assembled InAs/GaAs QDs (few per $\rm \mu m^{2}$), grown at the center of a planar optical microcavity by molecular-beam epitaxy. The microcavity consists of two distributed Bragg reflectors (15.5 lower and 10 upper AlAs/GaAs quarter-wave pairs) and a full-wave GaAs spacer. All the experiments described here were realized in a liquid He bath cryostat at a constant base temperature near 4.2 K, at which the QD ensemble emission spectrum spans $\approx$880 to 990 nm. The cavity mode of interest is located at $\approx$920 nm and coupled the QD emission to a 0.6-numerical aperture fiber-coupled microscope objective (probe area $ \approx$1 $\rm\mu m^2$). This objective was also used for above-band excitation with either a 633 nm HeNe laser or a ps pulsed mode-locked Ti:sapphire laser at 730 nm. In contrast, the near-resonant control laser, a tunable cw Ti:sapphire ring laser, was introduced laterally in the waveguide mode of the cavity so as to minimize stray laser scattering into the collection mode \cite{muller2007rfc}. This was achieved by bonding a single mode fiber directly to the cleaved edge ($[\bar{1}10]$) of the sample resulting in a compact and durable package. Optical measurements relied on a scanning Fabry-P\'erot (FP) interferometer with a free spectral range of 17 GHz and resolution  of $\approx$150 MHz, in series with a grating spectrometer.

\begin{figure*}[t!]
\includegraphics[width=6.7in]{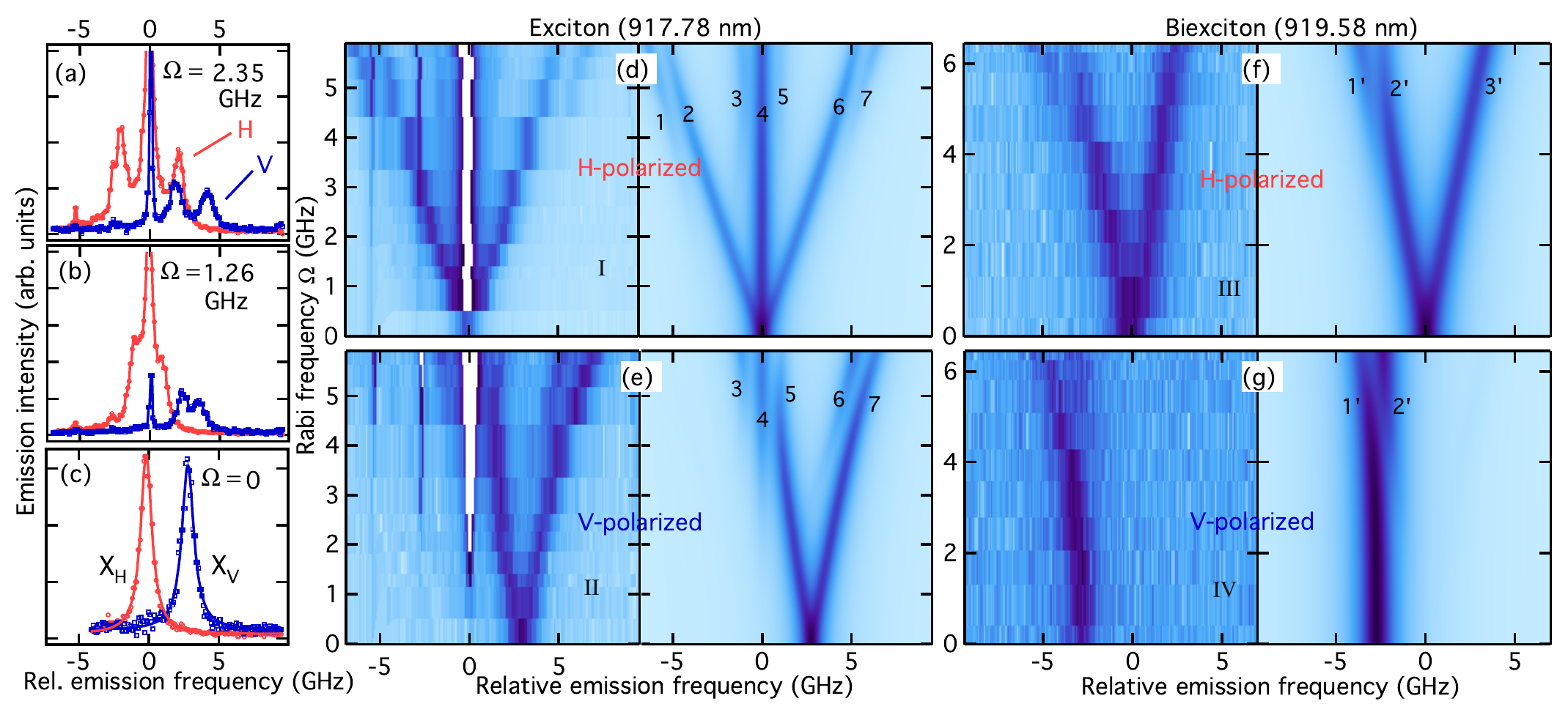}
\caption{\label{fig2:wide} (color online). Splittings of exciton and biexciton emission lines with variable Rabi frequency. (a-c) show exciton emission spectra at specific Rabi frequencies. (d-g) show full color-scale maps of emission intensity vs. emission (abscissas) and Rabi (ordinates) frequency. Panels to right represent simulated data. Roman numerals refer to the boxed transitions in Fig. 1. In the simulation panels, transitions are numbered, where primes refer to the biexciton transitions. In (d) and (e), the clear regions around the laser frequency correspond to signal truncation due to excessive background laser scattering.}
\end{figure*}

With such a combination we resolve the full spectrum of a single QD, particularly its fine-structure. For the neutral ground state excitons, the well-known fine-structure consists of a linear polarization doublet ($\rm |X_H\rangle$, $\rm|X_V\rangle $) split by $\delta_{\rm HV}$ due to the anisotropic exchange interaction \cite{gammon1996fss, bayer2002fsn}. ``H'' and ``V'' denote the two orthogonal in-plane polarization directions of these states. While we measure the relative orientation between polarizations we cannot determine their absolute orientation to the sample because of our fiber-based system. The neutral bound exciton pair ($\rm|XX\rangle$), the biexciton, has a two-path radiative cascade to the exciton vacuum ($\rm|0\rangle$) through the two possible $\rm|X_{H/V}\rangle$ states [Fig. 1(a)]. Each path produces photon pairs of identical polarization $\rm XX_H/X_H$ and $\rm XX_V/X_V$ with an additional energy difference between $\rm XX_{H/V}$ and $\rm X_{H/V}$ due to the biexciton binding energy, here $E_{\rm B}/\hbar\approx$650 GHz (2.6 meV) \cite{akopian2006epp}. For the QD studied, we measured excitonic linewidths of $\approx$1 GHz (4 $\mu$eV) and a splitting of  $\delta_{\rm HV}$=2.75 GHz (11 $\rm \mu$eV). The exciton and biexciton emission was identified on the basis of intensity-dependent PL measurements in which a square law dependence was observed for the biexcitonic emission.

We now apply the resonant control laser nominally polarized along $[\bar{1}10]$. We initially assume that the laser polarization at the QD is along H so that it couples only to $\rm X_H$. Then, the unperturbed states $\rm|0\rangle \otimes$$|N+1\rangle$ and $\rm|X_H\rangle$$\otimes |N\rangle$ are dressed into pairs of hybrid atom/field states $|1(N)\rangle$ and $|2(N)\rangle$ \cite{cohentannoudji}. $N$ represents the photon number of the applied field of frequency $\omega$. The splitting between the two states equals $\Omega'=\sqrt{\Omega^2+\delta^2}$ in the presence of laser detuning $\delta=\omega-\omega_0$; $\omega_0$ is the resonance frequency of the unperturbed transition, and $\Omega$ is the Rabi coupling of the resonant laser to $\rm X_H$. Transitions between adjacent photon number manifolds give rise to a resonant emission spectrum consisting of a central line at $\omega$ and side-bands at $\omega \pm \Omega'$. This is illustrated in Fig. 1(b), where the control laser, originally resonant with $\rm X_H$ creates four H-polarized transitions (two of which are degenerate), represented in the box labeled ``I''. In addition, Autler-Townes type splittings are expected from all transitions that share a common state with the states to which the laser couples. In Fig. 1(b) these are labeled ``II'' (V-polarized) and ``III'' (H-polarized).

Experimental results as a function of Rabi frequency are shown in Fig. 2. Detailed exciton emission lineshapes are displayed in Figs. 2(a-c) at $\Omega$=(0, 1.26, 2.35) GHz, respectively, for both H and V polarizations. Note that the FP frequency (abscissas) is relative to the frequency of $\rm X_H$ in the excitonic transitions and relative to the frequency of $\rm XX_H$ in the biexcitonic transitions. As the Rabi frequency increases (by increasing the control laser intensity), $\rm X_H$ indeed splits into a triplet [Fig. 2(d)] according to the illustration in Fig. 1(b). In Fig. 2(e,f) $\rm X_V$ and $\rm XX_H$ split into doublets. The remaining biexciton transition, $\rm XX_V$, is unaffected at low Rabi frequency [Fig. 2(g)] as anticipated [IV in Fig. 1(b)].

At larger Rabi frequencies the observations deviate from the simple picture given in Fig. 1(b). Additional features appear, that are, nonetheless, well-described by including some coupling of the laser to $\rm X_V$. A full simulation based on a Hamiltonian with an interaction term written as
 $H_{int}=-\frac{\hbar}{2}\sum_{j={\rm H,V}}{\Omega_j|{\rm X}_j,N\rangle\langle0,N+1|}/\sqrt{N}+h.c.$
can describe the results very well. For this particular QD, $\Omega_{\rm V}/\Omega_{\rm H}$=0.28. Since the coupling is related to the inner product of the field and optical dipole, either H is not aligned with $[\bar{1}10]$, the control field is significantly altered in the waveguide, or both. In this case we obtain 9 transitions (three dressed states in the $N$ manifold connecting the three states in the $N-1$ manifold), and out of those, 7 are nondegenerate (explicitly labeled in the Fig. 2 simulation panels). This is represented in Fig. 1(c) along with the three Autler-Townes transitions. The different spacings between dressed states reflect the different couplings and detunings to the two excitonic transitions.

\begin{figure}[h!]
\includegraphics[width=3.1in]{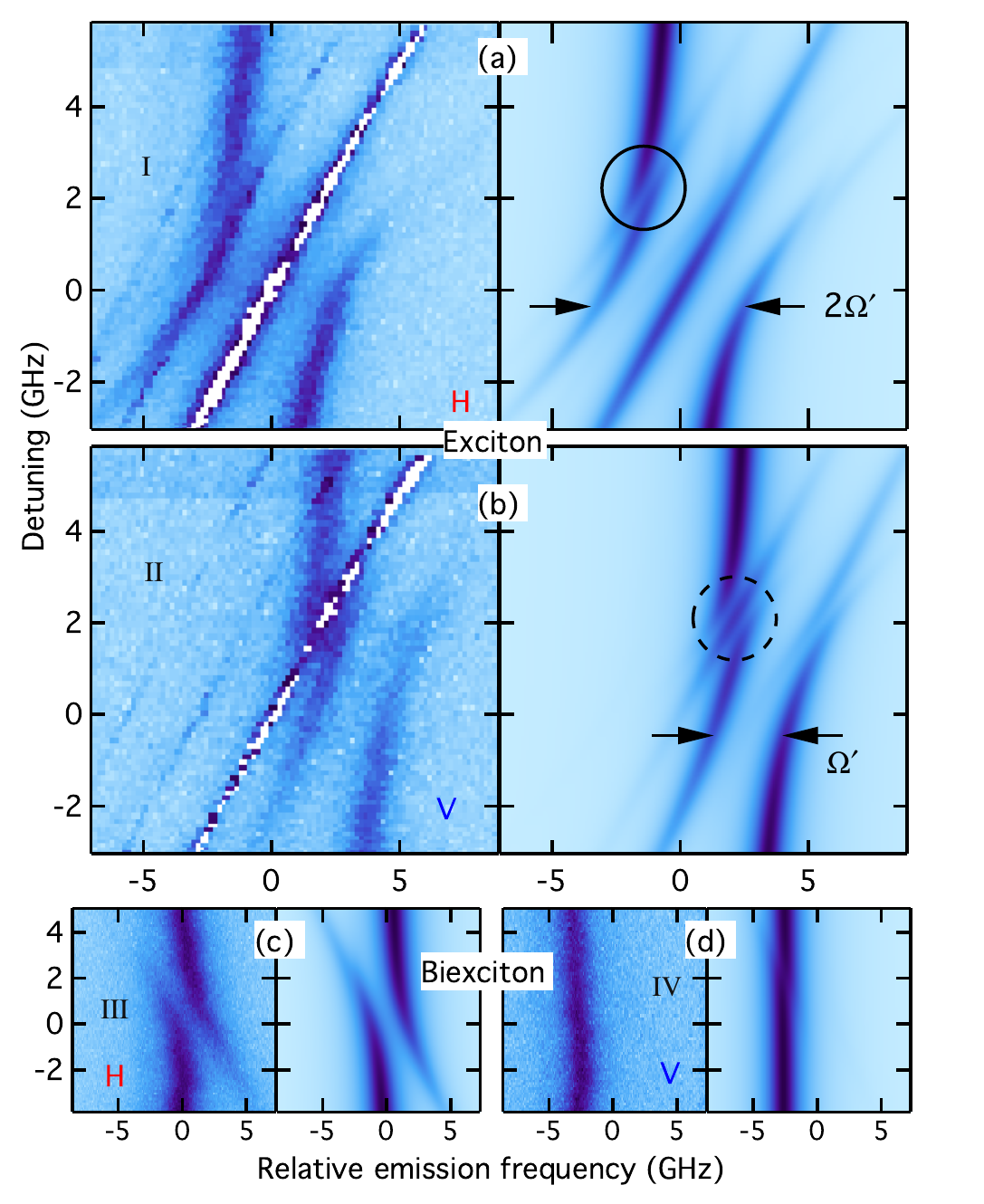}
\caption{\label{fig3} (color online). Emission spectrum of exciton (a,b) and biexciton (c,d) transitions under detuning for a fixed Rabi frequency of 2.4 GHz. Panels to right represent simulated data.}
\end{figure}

The dressed states undergo avoided crossings with laser detuning, shown in Fig. 3 for a fixed Rabi frequency of 2.4 GHz. Fig. 3(a) shows the direct transitions (I), in which the avoided crossing separates the dressed states by an amount ($2\Omega'$), with a minimum equal to 2$\Omega$ at resonance. The Autler-Townes transitions (II and III) are shown in Fig. 3(b,c) and have splittings equal to $\Omega'$. At large detuning, the optically active transitions asymptotically approach the original excitonic transitions, where the remaining finite energy offset is the AC Stark or light shift. The panels on the right in Fig. 3 show the result of simulations in excellent agreement with the data. There, the linewidths were held constant and light shifts of $\rm |XX\rangle$ due to the laser near $\rm X_{H/V}$ were neglected. This is very reasonable because the biexcitonic binding energy is much larger than $\hbar\delta_{\rm HV}$ and $\hbar\Omega$. As in the resonant case (Fig. 2) there are additional features due to the partial $\rm X_V$ coupling. For instance, when the laser is resonant with $\rm X_V$ then we observe a small triplet on $\rm X_V$ (dashed circle) and an Autler-Townes doublet on $\rm X_H$ (full circle).

For convenience the measurements above were performed with a HeNe laser as an excitation source. However, as shown in Fig. 4, the above-band excitation can be pulsed as well, giving rise to triggered single-photon emission for correlation measurements and on-demand sources. To simultaneously illustrate the capability to drive the biexciton rather than the exciton, the resonant laser is now applied to the $\rm XX_H$  transition (with the pump still on). The H-polarized emission from transitions to the ground state is then sent to a Hanbury-Brown-Twiss setup for second order correlation. The FP-resolved spectrum appears as expected as an Autler-Townes doublet but since the pump is now pulsed the second order correlation function $g^{2}(\tau)$ exhibits peaks spaced by the laser repetition period (13 ns) with a nearly missing peak at $\tau=0$ due to photon anti-bunching \cite{michler2000qds}. This remnant peak is unaffected by the presence of the control laser within the uncertainty of our measurements. We have also verified that the dressing did not depend on the pump laser. But as expected, when the pump laser intensity increases, the emission increases until saturation occurs.

In the absence of the above-band pump laser, we would expect the resonant control laser to generate resonance fluorescence from the resonantly-dressed transitions in region I, Fig. 1. However, without the pump laser, no such emission could be clearly identified. In our current setup, the emission at the laser frequency (within our FP bandwidth) was still dominated by indirect laser scattering so that in the elastic (coherent) regime the resonance fluorescence would go unnoticed. On the other hand we cannot exclude the presence of spectral shifts observed in previous absorption experiments \cite{xu2007cos, jundt2008ode}, due apparently to the strong resonant laser, that may detune and thus strongly weaken the resonance fluorescence.

\begin{figure}[t!]
\includegraphics[width=3.5in]{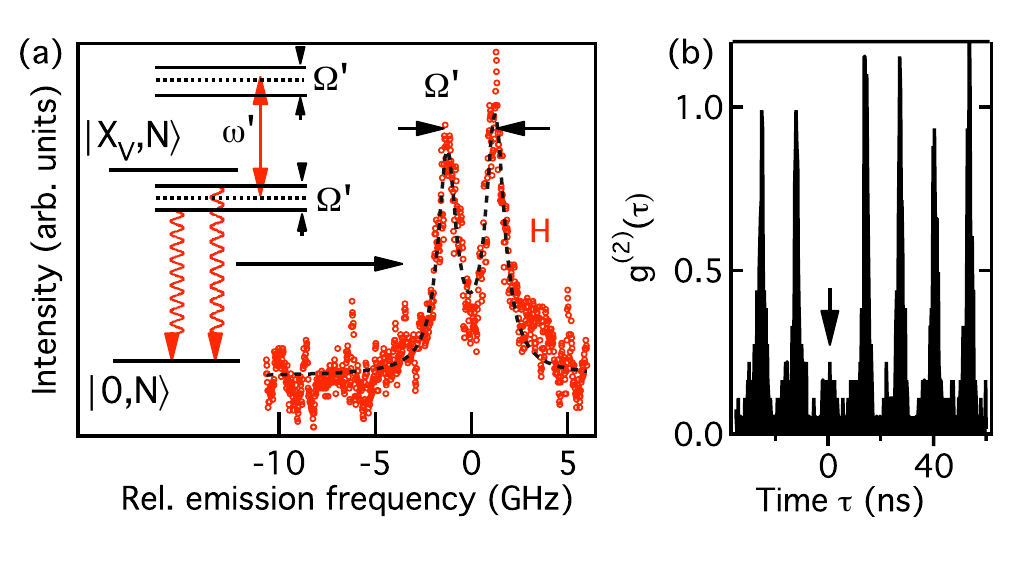}
\caption{\label{fig4} (color online). Autler-Townes doublet in the H-polarized exciton emission spectrum (a) and the second-order correlation function (b) of the PL of the exciton emission under resonant dressing of $\rm XX_{H}$.  The PL uses a 76 MHz (13 ns) pulsed laser at 780 nm, and the near absence of a peak at $\tau$ = 0 indicates single photon emission from the exciton state.}
\end{figure}

The drawback of the current approach mainly lies in the lack of control over the excitation polarization due to the fixed side excitation, which can be overcome with additional technical complexity. To excite an arbitrary superposition of $\rm X_V$ and $\rm X_H$ one could for example angle-polish the sample or introduce a second phase-locked laser into the wave-guide, rotated by 90$^{\circ}$ to the first one. Once the coupling to individual exciton or biexciton states can be controlled, the fine-structure splitting can be removed and deterministically-entangled photon emission established.

In conclusion, the complete biexciton-exciton emission spectrum of a single QD dressed by a strong resonant laser field was recorded for the first time. Combining a scanning FP with a microcavity waveguide excitation geometry that minimizes background laser scattering was found to be sufficient to observe all the salient features of the emission from this system, including resonant triplets and Autler-Townes doublets. This opens up avenues for utilizing the dressed system in new experiments and applications such as single and entangled photon sources, cavity-coupled QDs, coherent population trapping and qubit photon interfaces.


\begin{thebibliography}{99}


\bibitem{bayer2000hse}{M. Bayer {\it et al.}, Nature {\bf 405}, 923 (2000).}
\bibitem{borri2001udt}{P. Borri {\it et al.}, Phys. Rev. Lett. {\bf 87},157401 (2001).}
\bibitem{xu2007cos}{X. Xu {\it et al.}, Science {\bf 317}, 929 (2007).}
\bibitem{jundt2008ode} {G. Jundt {\it et al.}, Phys. Rev. Lett. {\bf 100}, 177401 (2008).}
\bibitem{muller2007rfc} {A. Muller {\it et al.}, Phys. Rev. Lett. {\bf 99}, 187402 (2007).}
\bibitem{wrigge2008ecp} {G. Wrigge {\it et al.}, Nat. Physics {\bf 4}, 60 (2008).}
\bibitem{cohentannoudji}{C. Cohen-Tannoudji, J. Dupont-Roc, and
  G. Grynberg, {\it Atom-Photon Interactions} (John Wiley and Sons, New York, 1992.)}
\bibitem{kroner2008rsa} {M. Kroner {\it et al.}, Appl. Phys. Lett. {\bf 92}, 031108 (2008).}

\bibitem{muller2007cds} {A. Muller {\it et al.}, arXiv:0707.3808v1 (2007).} 

\bibitem{unold2004ose} {T. Unold {\it et al.}, Phys. Rev. Lett. {\bf 92}, 157401 (2004).} 

\bibitem{bonadeo1998coc}{N. H. Bonadeo {\it et al.}, Science {\bf 282}, 1473 (1998).}

\bibitem{stievater2001roe}{T. H. Stievater {\it et al.}, Phys. Rev. Lett. {\bf 87}, 133603 (2001).}

\bibitem{htoon2002iro}{H. Htoon {\it et al.}, Phys. Rev. Lett. {\bf 88}, 087401 (2002).}

\bibitem{kamada2001ero}{H. Kamada {\it et al.}, Phys. Rev. Lett. {\bf 87}, 246401 (2001).}

\bibitem{zrenner2002cpt}{A. Zrenner {\it et al.}, Nature {\bf 418}, 612 (2002).}

\bibitem{michler2000qds}{P. Michler {\it et al.}, Science {\bf 290}, 2282 (2000).}


\bibitem{santori2001tsp}{C. Santori {\it et al.}, Phys. Rev. Lett. {\bf 86}, 1502 (2001).}

\bibitem{akopian2006epp}{N. Akopian {\it et al.}, Phys. Rev. Lett. {\bf 96}, 130501 (2006).}

\bibitem{stevenson2006sst}{R. M. Stevenson {\it et al.}, Nature {\bf 439}, 179 (2006).}

\bibitem{deppe1999ese}{D. G. Deppe {\it et al.}, IEEE J. Quantum Electron. {\bf 35}, 1502 (1999).}


\bibitem{gerard1998ese}{J.-M. G\'erard {\it et al.}, Phys. Rev. Lett. {\bf 81}, 1110 (1998).}

\bibitem{yoshie2004vrs}{T. Yoshie {\it et al.}, Nature {\bf 432}, 200 (2004).}

\bibitem{reithmaier2004scs}{J. P. Reithmaier {\it et al.}, Nature {\bf 432}, 197 (2004).}

\bibitem{peter2005eps}{E. Peter {\it et al.}, Phys. Rev. Lett. {\bf 95}, 067401 (2005).}

\bibitem{park2006cqd}{Y. S. Park, A. K. Cook, and H. Wang, Nano Lett. {\bf 6}, 2075 (2006).}



\bibitem{santori2006cpt}{C. Santori {\it et al.}, Phys. Rev. Lett. {\bf 97}, 247401 (2006).}

\bibitem{yao2005tcs}{Wang Yao and Ren-Bao Liu and L. J. Sham, Phys. Rev. Lett. {\bf 95}, 030504 (2005).}


\bibitem{kiraz2004qds}{A. Kiraz, M. Atat{\"u}re, and A. Imamoglu, Phys. Rev. A {\bf 69}, 032305 (2004).}

\bibitem{gammon1996fss}{D. Gammon {\it et al.}, Phys. Rev. Lett. {\bf 76}, 3005 (1996).}

\bibitem{bayer2002fsn}{M. Bayer {\it et al.}, Phys. Rev. B {\bf 65}, 195315 (2002).}


\end{thebibliography}
\end{document}